\pdfoutput=1
\documentclass[sigconf, screen]{acmart}
\usepackage{graphicx} % Required for inserting images
\usepackage{hyperref}
\usepackage[nameinlink]{cleveref}
\usepackage{booktabs}
\usepackage{tikz}
\usetikzlibrary{positioning}
\usetikzlibrary {graphs.standard}
\usepackage{pgfplots}
\usepackage{microtype}

\setcopyright{rightsretained}
\copyrightyear{2024}
\acmYear{2024}
\acmDOI{10.1145/3663533.3664041}
\acmSubmissionID{fsews24promisemain-id17-p}

\acmConference[PROMISE '24]{Proceedings of the 20th International Conference on Predictive Models and Data Analytics in Software Engineering}{July 16, 2024}{Porto de Galinhas, Brazil}
\acmBooktitle{Proceedings of the 20th International Conference on Predictive Models and Data Analytics in Software Engineering (PROMISE '24), July 16, 2024, Porto de Galinhas, Brazil}

\received{2024-03-28}
\received[accepted]{2024-04-19}
  
\acmISBN{979-8-4007-0675-2/24/07}



\begin{document}

\title{Prioritising GitHub Priority Labels}

\author{James Caddy}
\orcid{0000-0002-6254-8982}
\affiliation{%
  \institution{University of Adelaide}
  \city{Adelaide}
  \country{Australia}
}
\email{james.caddy@adelaide.edu.au}

\author{Christoph Treude}
\orcid{0000-0002-6919-2149}
\affiliation{%
  \institution{Singapore Management University}
  \city{Singapore}
  \country{Singapore}
}
\email{ctreude@smu.edu.sg}

\begin{abstract}
% Provide a brief summary of the paper, including the purpose of the data set, its key characteristics, and potential applications. Highlight the contributions and significance of the data set.
Communities on GitHub often use issue labels as a way of triaging issues by assigning them priority ratings based on how urgently they should be addressed. The labels used are determined by the repository contributors and not standardised by GitHub. This makes it difficult for priority-related reasoning across repositories for both researchers and contributors. Previous work shows interest in how issues are labelled and what the consequences for those labels are. For instance, some previous work has used clustering models and natural language processing to categorise labels without a particular emphasis on priority. With this publication, we introduce a unique data set of 812 manually categorised labels pertaining to priority; normalised and ranked as low-, medium-, or high-priority. To provide an example of how this data set could be used, we have created a tool for GitHub contributors that will create a list of the highest priority issues from the repositories to which they contribute. We have released the data set and the tool for anyone to use on Zenodo because we hope that this will help the open source community address high-priority issues more effectively and inspire other uses.
\end{abstract}

\maketitle

\section{Introduction}
% Clearly state the problem or gap in the existing research that the data set addresses.
GitHub is a collaborative tool used by hundreds of millions of users worldwide. One feature of GitHub is the ability to label issues within a repository's issue tracker with free-form text. This provides a non-opinionated method for contributors to categorise certain issues as they want. When problems arise during development of the repository or suggestions to improve the repository are added to the issue tracker as issues, there can be hundreds or thousands of issues that need to be prioritised for action. This leads GitHub repository maintainers to create their own system of labelling to sort the priorities of issues.

% Provide background information on the relevance of the data set to current software engineering challenges or trends.
In the last 20 years, software engineering research has been interested in the data available in bug-tracking systems such as Bugzilla, Jira, GitHub Issues, and others~\cite{canfora2005impact,fischer2003analyzing}. Many believe that this information can inform software process improvement, providing greater insight into the development process~\cite{mejia2017data,baysal2013informing,sureka2015ahaan}. Presently, GitHub is drawing more users than other bug-tracking systems that use structured or pre-defined priority labelling systems. This poses an increasing problem with researchers using contemporary issue priority data as a source of information, since more and more issue priority data is being labelled with free-form input on GitHub.

% Present the objectives of the data set creation and describe its intended use in the software engineering community.
A standardised set of priority labels would be helpful to the software engineering and research community, as a standard across repositories allows priority-related reasoning between repositories. For developers of multiple repositories, or newcomers to a repository, standardisation provides context to aid these developers in choosing their impact when contributing. Tooling that feeds data back to GitHub, or forward to third-party services, can benefit from standardised priority-labelling data to allow unified views, reporting, and interoperability.

It is for these reasons that we have created a data set that categorises labels from many popular repositories on GitHub into a standardised priority ordering. This data is made available for all in the hope that it can be used by people who will use it as intended above or will find their own applications of the data to contribute to the software engineering community.

\section{Related Work}
% Review existing literature and data sets related to the research topic.
In related work, authors sometimes identify labels that are priority-related, but do not release any data sets documenting these classifications.
% In related work, numerous authors identify that many labels are priority-related, but do not release any data sets documenting these classifications.
In one of the very few papers that has released a mapping of labels to priorities, Izadi, Akbari \& Heydarnoori~\cite{izadi2022predicting} list 33 high-priority and 14 low-priority labels from 70 repositories.

Li \emph{et al.} discuss training a deep learning neural network with priority-related labels in their paper on priority prediction~\cite{li2022atale}. They identify multiple projects using such labels, give a few examples of high-priority labels, and discuss the normalisation of priorities into three categories; the same ones we have used in this paper.

Kim and Lee in their paper \emph{An Empirical Study on Using Multi-Labels for Issues in GitHub} express frustration and state that custom labels ``might be one of the huge obstacles that interfere[s] with label analysis on issue management''~\cite{kim2021empirical}. Although the two authors describe some labels as ``issue priority'' labels, they have not provided a list of labels and corresponding categories.
% and give a couple of examples, in the data that they have released, they do not provide a list of issues and corresponding categories.

In the paper titled \emph{Exploring the use of labels to categorize issues in Open-Source Software projects}~\cite{cabot2015exploring}, the authors, Cabot \emph{et al.}, describe four typical labelling strategies in issue trackers: priority labels, versioning labels, workflow labels, and architectural labels. These strategies were discovered through a clustering algorithm, but the clusters and the labels within have not been released as public data to reference.

% Highlight the limitations or gaps in the current data sets that the new data set aims to address.
% Discuss how the data set complements or extends existing resources.
This data set represents a hand-categorised set of issues, where most existing work has used machine learning in the form of clustering models~\cite{cabot2015exploring} or natural language processing~\cite{diniz2020github} to achieve a similar categorisation.
% Without any publicly available data set of machine learning-generated mappings, it is impossible to determine the accuracy of the categorisations such as those that have already been used in study methods.

Automatic categorisation is subject to problems just as hand categorisation is, but the two sets complement each other with the nuance of human rating~\cite[\S5.A]{kim2021empirical}, and the thoroughness of automatic rating. Training and validation typically relies on hand-categorised data, and larger data sets can contribute to higher accuracy machine learning models. This data set also represents one of the very few publicly available mappings of issue labels to priority ratings.

\section{Data Description}
% Provide a detailed and comprehensive description of the data set.
Using the GitHub API we retrieved the labels from the the top 5,000 most `starred' repositories on GitHub.
% We used the GitHub API to retrieve all the labels used in the top 5,000 most `starred' repositories on GitHub.
This decision was based on efficiently maximising the number of people who use the labels that are categorised. New projects, commercial projects and the wider software engineering community might not see this data set generalise to their applications.
% Do note that this is limiting of the generalisability to newly created projects, mature projects on other platforms, commercial projects and a the wider software engineering community.
Of the resulting 29,168 labels, we categorised 812 of those as being related to the priority of the issue. These 812 labels are then ranked into three levels of priority, `High', `Medium', and `Low'. We have chosen this scale to represent a distinctive set of priorities, as Li \emph{et al.} do~\cite{li2022atale}. 

Approximately 1,667 repositories use custom priority labels. 897 use at least one high-priority label, 339 use at least one medium-priority label, and 1,295 use at least one custom low-priority label. We have elected to only count those that have used custom low-priority labels, as GitHub provides the ``wontfix'' label by default with repositories that have not customised their labels, leading to an over-representation.

A table of the most-used priority labels, and how many repositories use them can be found in \Cref{tab:prioritylabelfrequency}. Although GitHub differentiates labels with different letter case, we have chosen not to recognise this distinction and have combined labels that differ only by letter case to show a broader selection of labels.

\begin{table}[htbp]
	\centering
	\caption{10 Most Adopted Priority Labels}
	\label{tab:prioritylabelfrequency}
	{
		\begin{tabular}{c|c}
            ``low'' & 531 \\
            ``priority'' & 406 \\
            ``high'' & 382 \\
            ``stale'' & 379 \\
            ``medium'' & 210 \\
            ``critical'' & 172 \\
            ``major'' & 167 \\
            ``future'' & 133 \\
            ``high prio'' & 128 \\
            ``minor'' & 127 \\
		\end{tabular}
	}
\end{table}

% Include information on the source of data, data collection methods, and any preprocessing steps applied.
GitHub provides label data through its REST API~\cite{githubrepolabels}. Through this API, each of the 5,000 most starred repositories can be queried for their labels. Label data is returned as JSON which contains the `name' field, which can then be extracted to the file containing all the labels for every repository queried. Beside the extraction from the JSON returned by the GitHub API, the data has not been pre-processed in any way.

% Clearly define the variables and features present in the data set.
The data set is formatted as a comma-separated value list of UTF-8 strings, representing the labels directly from GitHub's API and the category of priority it is assigned. The categories are the three levels defined previously; `High', `Medium', and `Low'.

Examples of labels that were assigned `Low' priority included `Inactive', `Won't Fix', or `Stale'. These examples are the most common types of labels that were not specifically referencing priority but still expressed that the development on the issue has stopped or was not needed and therefore low priority to complete. Examples of labels that were assigned `High' priority included `important', or `urgent'. 

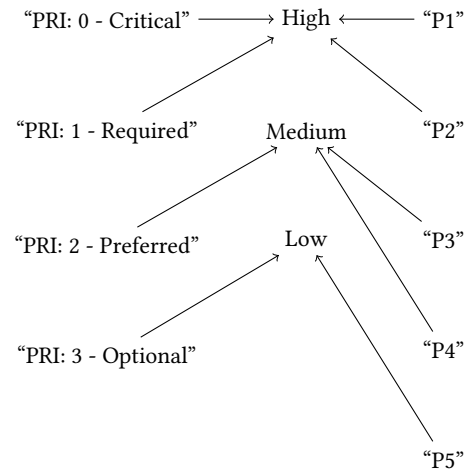
\begin{figure}[htbp]
    \centering
    \caption{Normalised Mapping of Different Scales}
    \label{fig:labelmapping}
    {
        \begin{tikzpicture}
            % Nodes for the first group of labels
            \node (high) {High};
            \node[below=of high] (medium) {Medium};
            \node[below=of medium] (low) {Low};
        
            % Nodes for the second group of labels
            \node[left=of high] (critical) {``PRI: 0 - Critical''};
            \node[below=of critical] (required) {``PRI: 1 - Required''};
            \node[below=of required] (preferred) {``PRI: 2 - Preferred''};
            \node[below=of preferred] (optional) {``PRI: 3 - Optional''};
        
            % Nodes for the third group of labels
            \node[right=of high] (P1) {``P1''};
            \node[below=of P1] (P2) {``P2''};
            \node[below=of P2] (P3) {``P3''};
            \node[below=of P3] (P4) {``P4''};
            \node[below=of P4] (P5) {``P5''};
        
            % Arrows connecting the nodes
            \draw[<-]
                  (high) edge (critical)
                  (high) edge (required)
                  (medium) edge (preferred)
                  (low) edge (optional)
                  (high) edge (P1)
                  (high) edge (P2)
                  (medium) edge (P3)
                  (medium) edge (P4)
                  (low) edge (P5);
        \end{tikzpicture}
    }
\end{figure}

An important characteristic of the ranking process was the normalisation of different scales. Many labels take the form of a numerical scale for priority, with one example of a scale being `PRI: 0 - Critical', `PRI: 1 - Required', `PRI: 2 - Preferred', and `PRI: 3 - Optional'. With four degrees of priority represented in this scale and only three of our own, we elect to break ties in favour of having higher priority. A visual example has been given in \Cref{fig:labelmapping}. Sometimes this required more context and information than just the label text gives, an example being the scale of `P5', ... `P1', where the ambiguity could mean that either end of the scale was the highest priority. To determine this information, the repository that uses the scale was searched for issues with the labels in question, and a subjective judgement based on the severity of the language used in the description and comments was made. 

% Discuss any challenges or biases in the data and how they were addressed.
There are some limitations to the data set that should be known. Firstly, the data was captured on the day of 2022-06-01, and the labels analysed only represent those that were present at the time. 

It is impossible to determine an inter-repository ranking of priority due to the subjectivity and incomparable nature of different repositories. Even with the abstractions afforded by the labelling systems that repositories use, it is not realistic to agree on different repositories having equivalent rankings of priorities. This is even the case when two different repositories share the same labels.

% GitHub is not restricted to English language users. 
Unfortunately, the authors are unable to accurately rank labels in languages other than English. For this reason, we have decided not to rank labels that are not English language.

The ranking of priority between labels is the opinion of the primary author. To check that these judgements were reasonable, we took a sample of 2,000 labels, and three participants were asked to assign priorities to the labels. We chose 2,000 as a representative sample of 29,168 to achieve a minimum 99\% confidence level and 1\% margin of error with the estimated proportion at 3\%. The participants were asked to give each label one of four categories, the three degrees of priority, and the fourth `conveys no priority' category. 

Once the authors had ranked the sample of issues, three rater agreement tests were performed on the data, the results of which can be seen in \Cref{tab:cohen'SUnweightedKappa}, \Cref{tab:fleiss'Kappa}, and \Cref{tab:krippendorff'SAlpha}. The results show good agreement between the raters\textendash despite some disagreements. The Fleiss' kappa analysis shows a majority of this disagreement comes from the medium and low-priority labels. The following labels are three identified as `low-priority' by the first rater but not the second or third: ``frozen-due-to-age'', ``wont do'', ``Status: stale''. The second rater identified these two `high-priority' labels that the first and third rater did not: ``big-bug'', ``daily blocker''. This shows a large amount of subjectivity in the judgement, but the raters strongly agree on most labels. Both kappa and alpha values being closer to $1.0$ show greater agreement. While there is no objective interpretation of these consensus values, several papers suggest that $0.68$ shows moderate agreement, below strong agreement~\cite{landis1977measurement,munoz1997interpretation,mchugh2012interrater}. For the purpose of this study, the agreement was deemed sufficient to allow one rater the task of ranking the remaining labels. To ensure that coverage was thorough, textual search of each label related to priority was undertaken, with case insensitivity and with\textemdash where applicable\textemdash``fragments'' of the label, e.g. word stems or single words of multi-word labels.

\begin{table}[htbp]
	\centering
	\caption{Cohen's Unweighted kappa}
	\label{tab:cohen'SUnweightedKappa}
	{
		\begin{tabular}{lrrrr}
			\toprule
			\multicolumn{1}{c}{} & \multicolumn{1}{c}{} & \multicolumn{1}{c}{} & \multicolumn{2}{c}{95\% CI*} \\
			\cline{4-5}
			Ratings & Unw. kappa & SE** & Lower & Upper  \\
			% Ratings & Kappa & SE** & Lower & Upper  \\
			\cmidrule[0.4pt]{1-5}
			Average kappa & $0.680$ & & &  \\
			Rating A - Rating B & $0.696$ & $0.060$ & $0.577$ & $0.814$  \\
			Rating A - Rating C & $0.683$ & $0.067$ & $0.552$ & $0.814$  \\
			Rating B - Rating C & $0.662$ & $0.068$ & $0.530$ & $0.795$  \\
			\bottomrule
			\addlinespace[1ex]
		\end{tabular}
	} \\
    \footnotesize{\textit{*~Confidence Interval. **~Standard Error.}}
\end{table}

\begin{table}[htbp]
	% \centering
	\caption{Fleiss' kappa}
	\label{tab:fleiss'Kappa}
	{
		\begin{tabular}{lrrrr}
			\toprule
			\multicolumn{1}{c}{} & \multicolumn{1}{c}{} & \multicolumn{1}{c}{} & \multicolumn{2}{c}{95\% CI*} \\
			\cline{4-5}
			Ratings & Fleiss' kappa & SE** & Lower & Upper  \\
			\cmidrule[0.4pt]{1-5}
			High & $0.779$ & $0.013$ & $0.754$ & $0.804$  \\
            Medium & $-0.001$ & $0.013$ & $-0.026$ & $0.024$ \\
			Low & $0.351$ & $0.013$ & $0.326$ & $0.376$  \\
			- & $0.695$ & $0.011$ & $0.670$ & $0.720$  \\
			Overall & $0.681$ & $0.011$ & $0.659$ & $0.702$  \\
			\bottomrule
			\addlinespace[1ex]
		\end{tabular}
	} \\
    \footnotesize{\textit{*~Confidence Interval. **~Standard Error.}}
\end{table}

\begin{table}[htbp]
	% \centering
	\caption{Krippendorff's alpha}
	\label{tab:krippendorff'SAlpha}
	{
		\begin{tabular}{lrrrr}
			\toprule
			\multicolumn{1}{c}{} & \multicolumn{1}{c}{} & \multicolumn{1}{c}{} & \multicolumn{2}{c}{95\% CI*} \\
			\cline{4-5}
			Method & Kripp. alpha & SE** & Lower & Upper  \\
			% Method & Alpha & SE** & Lower & Upper  \\
			\cmidrule[0.4pt]{1-5}
			Nominal & $0.681$ & $0.055$ & $0.568$ & $0.775$  \\
			\bottomrule
			\addlinespace[1ex]
		\end{tabular}
	} \\
    \footnotesize{\textit{*~Confidence Interval. **~Standard Error.}}
\end{table}

\section{Data Characteristics}
\begin{figure} [htbp]
    \centering
    \caption{Bar Chart of Rating Frequencies}
    \begin{tikzpicture}
        \label{fig:barChartFrequencies}
        \begin{axis} [
            ybar,
            symbolic x coords={Low, High, Medium},
            xtick=data,
            bar width=20pt,
            width=.45\textwidth,
            height=.3\textwidth
        ]
        \addplot coordinates {
            (High, 292)
            (Medium, 102)
            (Low, 417)
        };
        \end{axis}
    \end{tikzpicture}
\end{figure}
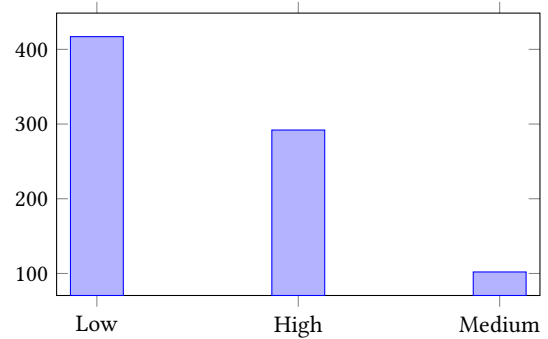

% Discuss key characteristics of the data set, such as size, complexity, and diversity.
The data set contains 29,168 labels, of which 812 were identified as priority-related. These 812 labels were classified into three categories of priority; 293 `high-priority', 102 `medium-priority', and 417 `low-priority'. The frequencies are recorded in \Cref{tab:frequenciesForRating}, and a visual representation of the distribution can be found in \Cref{fig:barChartFrequencies}.

% Include statistics and visualizations to help readers understand the distribution of data points.
% Highlight any unique or noteworthy aspects of the data set.

\begin{table}[htbp]
	\centering
	\caption{Frequencies for Rating}
	\label{tab:frequenciesForRating}
	{
		\begin{tabular}{lrrr}
			\toprule
			Rating & Frequency & Percent & Ranked Percent  \\
			\cmidrule[0.4pt]{1-4}
			High & $292$ & $1.001$ & $36.005$  \\
			Medium & $102$ & $0.350$ & $12.577$  \\
			Low & $417$ & $1.430$ & $51.418$  \\
			- & $28,357$ & $97.220$ & $ $  \\
			Total & $29,168$ & $100.000$ & $ $  \\
			\bottomrule
		\end{tabular}
	}
\end{table}

\section{Use Cases}
% Present practical examples or use cases where the data set can be applied in software engineering research or practice.
Consider an open-source software developer who contributes to many GitHub repositories. These repositories may use different labelling systems, making it difficult to to identify high-priority issues quickly. 
% There are multiple issues in those repositories that are awaiting resolution, but the developer needs to know which are the most important and that require immediate action.
% These different repositories could use very different triage processes, or none at all, which makes the work-discovery process more complicated.
We have created a python script with the data set that will take a GitHub user ID and return a prioritised list of issues from all the repositories they contribute to.
% We have created a tool to help alleviate this problem. Along with the data set, we are including a Python script that will take a GitHub user ID and return a list of issues in decreasing priority from all the repositories to which they have contributed. 
With this tool, the developer in the example would be able to see which issues from all their repositories are high-priority without having to remember which labels are part of a repository's process or search for all labels that users may have labelled their issues in a less structured system. This Python script is a basic example but may provide inspiration or a starting point for a more feature-rich tool. 

The creation of this data set was motivated by the need to identify priority-labelled issues for the registered report \emph{Is Surprisal in Issue Trackers Actionable?}~\cite{caddy2022surprisal}, which examimes the correlation between an issue being surprising and its assigned priority.
% The motivating example for creating this data set came from a need to identify priority-labelled issues for the registered report \emph{Is Surprisal in Issue Trackers Actionable?}~\cite{caddy2022surprisal}. In this paper, the authors sought to find out if there is a correlation between an issue being surprising and that issue being assigned a high-priority. For this, it was required that a high-quality categorisation of priority-labels was created for the repositories analysed.

% A large proportion of critical bugs that are left unresolved may be an indicator of the health of a software repository. Quality problems indicated by a large number of open high-priority issues can be redressed if a focused effort is made to correct the course of the project. This could be especially important for a company with several repositories and no software health monitoring or monitoring that does not take into account open issues and their severities.

% Discuss potential benefits and insights that researchers can gain from using the data set.
Better prediction surrounding ideas of issue priority could benefit matters of software engineering. Many researchers are interested in who should fix high-priority issues~\cite{anvik2006who}, how long it will take to fix high-priority issues~\cite{weiss2007how,bhattacharya2011bugfix,giger2010predicting}, how to predict the number and frequency of high-priority issues and the related automatic triage of high-priority issues~\cite{bugayenko2023prioritizing,izadi2022predicting,lamkanfi2010predicting}. 
% With a better understanding of which issues are high-priority, aided by a more complete mapping of free-form priority-related labelling, the predictive power of such prediction algorithms is increased.

A set of hand-ranked labels may serve as a set of annotated data for machine learning purposes. Some related work uses machine learning as a way of categorising issues, for instance, in the paper titled \emph{GitHub Label Embeddings} by Diniz \emph{et al.}~\cite{diniz2020github}.

The data set of Izadi, Akbari, and Heydarnoori~\cite{izadi2022predicting} that lists 47 priority-related issues has already been used by He \emph{et al.} in their paper \emph{Understanding and Enhancing Issue Prioritization in GitHub}~\cite{he2023understanding} that analyses the effectiveness of priority-labelled issues.

\section{Availability and Access}
% Specify how the software engineering community can access and use the data set.
The data set has been made accessible through Zenodo~\cite{caddy2024dataset}.
% Provide information on any restrictions, licenses, or terms of use.
It is licensed under the Creative Commons 4.0 Attribution license.

\section{Summary}
% Summarize the key contributions and findings of the paper.
GitHub repositories employ issue labels to prioritise tasks, yet due to their un-opinionated labelling system, a lack of standardisation impedes cross-repository reasoning for researchers and users alike. This paper introduces a novel data set of 812 GitHub issue labels categorised into low, medium, and high priorities. Additionally, a tool is provided to assist GitHub contributors in identifying high-priority issues across the repositories they contribute to. Both data set and tool are made publicly available, aiming to enhance the software community's efficiency in addressing high-priority issues.

% Discuss potential future work, including any planned updates or expansions to the data set.
By providing a categorised data set and tool to aid work prioritisation, we hope that software engineering practices and their research can be enhanced through additional applications that leverage this data set. The applicability can further be improved by adding depth\textemdash categorising more labels as priority-related\textemdash or by adding breadth\textemdash categorising labels according to categories matching the other principle labelling strategies that are commonly used on GitHub.

\bibliographystyle{ACM-Reference-Format}
\bibliography{bib}   % name your BibTeX data base

\end{document}